# Enhanced contact angle hysteresis of salt aqueous solution on graphite surface by a tiny amount of cation

Haijun Yang[1,3#], Yangjie Wang[3,6#], Yingying Huang[1], Shuai Wang[3,4], Qiufeng Lü[5], Xiaoling Lei[1,2], and Haiping Fang[1,2,3*]

1. Shanghai Synchrotron Radiation Facility, Zhangjiang Laboratory (SSRF, ZJLab), Shanghai Advanced Research Institute, Chinese Academy of Sciences, Shanghai 201204
2. Department of Physics, East China University of Science and Technology, Shanghai 200237, China
3. Division of Interfacial Water, CAS Key Laboratory of Interfacial Physics and Technology, Shanghai Institute of Applied Physics, Chinese Academy of Sciences, Shanghai 201800, China.
4. School of Environmental and Chemical Engineering, Shanghai University, Shanghai 200444, China
5. College of Materials Science and Engineering, Fuzhou University, Fuzhou 350116, China
6. University of Chinese Academy of Sciences, Beijing 100049, China

\# These authors contributed equally to this work
*Corresponding author. E-mail: fanghaiping@sinap.ac.cn (H.F.)

Abstract: We experimentally observed the enhanced contact angle hysteresis (CAH) of dilute aqueous salt solution on graphite surface, i.e., $40.6\pm0.6°$, $34.6\pm1.4°$, and $27.8\pm0.6°$, for LiCl, NaCl, and KCl, indicating the effective tuning of the CAHs by cations. Molecular dynamics simulations reveal that the preferential adsorption of cations on the HOPG surface due to the cation-π interaction pins the water at the backward liquid-gas-solid interfaces, reducing the receding contact angle and hence enhancing the CAH. This finding provides a simple method to control the contact angle and the CAH of aqueous drops on graphitic surfaces such as graphene, carbon nanotube, biomolecules, and airborne pollutants.

Contact angle hysteresis (CAH), omnipresent in the nature and various industrial processes, is one of the most important and classic elements of wetting of water droplets on a solid substrate and an outstanding challenge in interfacial dynamics [1-5]. The existence of CAH affects various physical phenomena relevant for industrial applications [6-8]. It has been reported that CAH can be caused by the surface roughness and chemical heterogeneities, surface deformation, chain length of the organic liquid molecule, and even the volume fraction of the microbead suspension, suggesting that the origin of CAH is complicated and far from full understood. [7-12] Enhancing CAH is critical to effective sticking a drop of ink/pesticide on a substrate, promoting the adhesion of biomaterials on surfaces, accelerating the speed of printing/coating, and improving the water-retention capacity of soil. [13-15] However, challenge still remains for simple, efficient, and environment-friendly control of the

CAH with less disturbance and low cost, which are key to those fundamental and industrial applications in a given system.

Aromatic-rich surfaces include graphitic materials such as graphene, carbon nanotube, graphite, and many polymers (polyethylene terephthalate, and polyimide), biomolecules, humus in soil, and even airborne pollutants. Particularly, the wetting behavior of graphene have attracted intense scientific interest since its discovery [16-25]. The wetting behavior of salt aqueous solution on these aromatic-rich surfaces is of great importance in the properties of permeation [26], adhesion [27,28], adsorption [29,30], hydrophilicity [31,32], and solubility [33], affecting their applications including cleaning [34], printing/coating [35], filtration [36-40], lubrication [41-44], battery and supercapacitors electrodes [45,46], digital microfluidics [47], heat transfer [48], catalyst and support [49,50], soil improvement [51,52], and air pollution control [53].

In this Letter, by combining experimental observations and theoretical calculations, we show that a tiny amount of cation ($K^+$, $Na^+$, and $Li^+$) can abruptly decrease the receding contact angle ($\theta_R$) of water droplets on a typical aromatic-rich material, highly oriented pyrolytic graphite (HOPG), and enhance the CAH. Molecular dynamics (MD) simulations and atomic-force-microscopy (AFM) images reveal that the preferential adsorption of cations on the HOPG surface due to the strong cation-π interaction pins the water at the backward liquid-gas-solid interfaces when the drop slides over, which reduces the receding contact angle but does not affect the advancing contact angle, resulting in an enhanced CAH. These findings provide a simple method to control the contact angle and the CAH of aqueous drops on graphitic surfaces such as graphene, carbon nanotube, graphite, humus, biomolecules, and airborne pollutants, shedding new light on improving related physical, chemical and biological processes.

The freshly cleaved HOPG substrate with a drop of salt aqueous solution sitting on its surface was attached to the hollow rotation stage between the light source and the camera, and slowly tilted by vertically rotating the stage until the drop just began to move (See Fig. S1). All contact angles were measured by a biolin Attension® Theta contact angle meter via the tilting plate method. A series of images were recorded by the camera, as a typical one shown in Fig. 1a, and their advancing ($\theta_A$) and receding ($\theta_R$) contact angles were analyzed with the built-in software. The experiment was repeated at least 8 times for each condition, and only the average and standard deviation of the series with the largest receding contact angles and the smallest advancing contact angles were chosen as $\theta_R$ and $\theta_A$, respectively.

As shown in Fig. 1b, $\theta_A$ and $\theta_R$ of the drop of pure water on HOPG are 71.0±0.3° and 60.3±0.1°, respectively. Their average value for pure water drop is 65.6±0.4°, very close to the previously reported value of 64.4° [22]. Interestingly, $\theta_R$ abruptly decreases at the initial stage with very low salt concentrations, and then maintains its average value with a fluctuation until the salt concentration reaches 0.80 M (mol/L). For example, $\theta_R$ decreases from its average of 60.3±0.1° for pure water to 38.9±0.8° at the salt concentration of 0.20 M for NaCl, and then fluctuates around this value. Moreover, $\theta_R$ shows the average values of 49.3±0.1° and 37.2±0.6° after the transition concentrations of 0.04 M for KCl, and 0.20 M for LiCl, respectively (Fig. S2). In contrast, $\theta_A$ increases only from 71.0±0.3° to about 75.1±1.4° at 0.10 M, 0.04 M, and 0.02 M for LiCl, NaCl, and KCl (Fig. 1b and S2), respectively, indicating that both $\theta_A$ and its corresponding salt transition concentration weakly depend on the kind of cations.

We calculated the contact angle hysteresis (CAH, $\Delta\theta=\theta_A-\theta_R$), for KCl, NaCl, and LiCl dilute aqueous solutions, as displayed in Fig. 1c. The CAH abruptly increases from 10.7±0.4° to 27.8±0.6°, 34.6±1.4°, and 40.6±0.6° when the salt concentrations arise from zero to no more than 0.40 M, 0.20 M, and 0.04 M for KCl, NaCl, and LiCl, respectively, then fluctuates around these values. That is, the CAH of dilute salt aqueous droplet reaches about 2.8 to 4.0 times of that of the pure water, indicating that the CAH is enhanced by the salts in these aqueous droplets. Even a tiny amount of salt is enough to enhance the CAH. Among them, KCl induces the lowest CAH at the highest transition concentration, while LiCl stimulates the highest CAH at the lowest transition concentration.

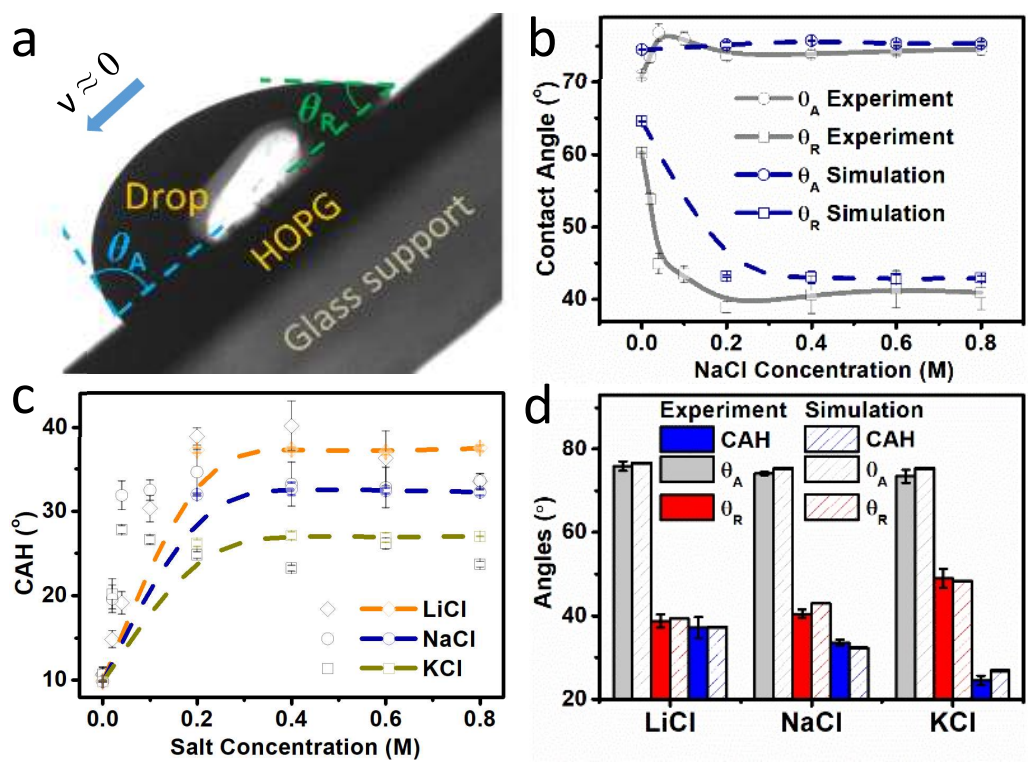

**Figure 1.** Contact angle hysteresis (CAH, $\Delta\theta=\theta_A-\theta_R$) of a dilute salt aqueous drop on HOPG. (a), A typical image of a dilute NaCl aqueous drop on a tilting HOPG substrate. $\theta_A$ and $\theta_R$ represent the advancing and receding contact angle at its leading and trailing edge, respectively. (b), Contact angle of a dilute NaCl aqueous drop on HOPG with the salt concentration of less than 1.00 M (mol/L). The solid and dashed spline lines denote the experiment and molecular dynamics (MD) simulation results, respectively. Error bars indicate the standard deviations. (c), Variation of CAH on the salts concentration.

Scattered symbols represent the experiment results, while the spline-fitted dash lines with the same symbols represent the MD simulation results. Error bars indicate the standard deviation. (d), The receding and advancing contact angles, and the average CAH (from 0.20 M to 0.80 M). Red, light grey, and blue bars represent the receding and advancing contact angles, and the CAH respectively. The solid bars represent experimental values while shadow ones denote the MD simulation results.

In order to illustrate the underlying mechanism, we performed MD simulations with a modification of the cation–π interaction [26] of salt aqueous drops on graphite. Considering the intrinsic atomic steps on the surface of HOPG in practice, the substrate consisted of a smooth graphene surface and graphite two-layer stripes at lateral spacing of 12.3 Å between the stripes on it in the simulation. We mimicked the gravitational effect to the salt aqueous drops on the tilted graphite surface in the experiments by applying an additional acceleration ($a$) along the positive $x$ direction of a striped graphene surface to a salt aqueous droplet. To quickly approach the critical contact angles during simulation, a dichotomy was adopted in the testing of the acceleration values for each droplet until the droplet just began to move. Fig. 2a displays a snapshot for a NaCl aqueous droplet for the critical acceleration $a = 5.1 \times 10^{-3}$ nm/ps$^2$. The values of $\theta_R$ and $\theta_A$ at different salt concentrations for a NaCl aqueous droplet are displayed in Fig. 1b and KCl and LiCl aqueous droplet in Fig. S2 in Supplementary Materials (SM). $\theta_R$ fluctuates around 37.0°, 43.0° and 50.0° for LiCl, NaCl, and KCl, respectively, while $\theta_A$ almost remains the value for the pure water droplet. As shown in Fig. 1c, the values of CAH fluctuates around 38.0°, 32.0°, and 25.0° for LiCl, NaCl, and KCl, consistent with the experimental observations. We attribute the enhancement of the CAH for saline droplets to the "pin" of the cations on the striped graphene substrate by strong cation–π interactions. We found that the average residence time of the cations in the region near the backward liquid-gas-solid interfaces is longer than 8 ns, i.e., in all the 8 ns MD simulations, these cations always stay in the region. These results confirmed that cations preferentially pinned on the substrate near the backward liquid-gas-solid interfaces. In order to further demonstrate the "pin" effect, we performed new MD simulations in which the cation–π interaction is removed from the force-field (see details in PS2 of SM), the CAH decreases to about 20° for NaCl droplet as the CAH experimentally obtained for the pure water droplet, and the average residence time of the cations near the backward liquid-gas-solid interfaces are only about 3.7 ns.

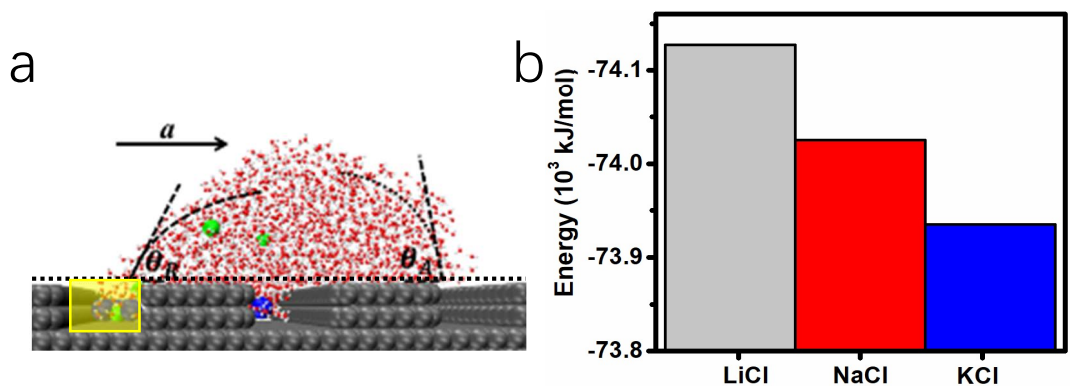

**Figure 2.** (a), Side-view snapshots of modeling system for an aqueous droplet on a graphene sheet after 8 ns MD simulation with a critical acceleration of $5.1 \times 10^{-3}$ nm/ps$^2$ for a NaCl droplet with cation-π interactions. Spheres in red and white represent oxygen and hydrogen atoms, respectively. Blue and green spheres represent cations (Li$^+$, Na$^+$, or K$^+$) and Cl$^-$, respectively. Black dotted lines represent the profile of the simulated water/vapor interface, fitted with two circles for the $\theta_R$ and $\theta_A$ calculation, respectively. The translucent yellow rectangle near the backward liquid-gas-solid interfaces illustrates the statistical region for cation staying. (b), Hydration energies (i.e. hydration capacity) of the modeling system comprising 2178 water molecules, one cation, and one chloride anion.

The enhancement of CAHs for saline droplets highly depends on the cation species. As displayed in Fig. 1c, the value of CAHs increases from about 9.9° for pure water to about 25.0°, 32.0°, 38.0° for KCl, NaCl and LiCl, respectively. We attribute the dependence of the enhanced CAH on the cation species to the different hydration capacity of cations pinned on the substrate. We found that the order from highest to lowest hydration capacity (i.e. energy) of cations on graphene with water molecules was Li$^+$ > Na$^+$ > K$^+$ in Fig. 2d, consistent with the order of the enhancement of CAH from the experimental observations in Fig. 1d.

We note that, if the substrate surface is perfectly smooth without atomic steps, cations will slip along the surface although there is a strong cation–π interaction between the cation and the surface. Thus, cations without pinning can hardly enhance the CAH. As shown in Fig. S4 in SM, the receding and advancing contact angles are equal on the surface without atomic steps. In reality, there are inherent atomic steps on HOPG surface, which provides "dams" to prevent cations from slipping. Thus, in the simulation, we set stripes on the graphene surface and expected that the stripes will result in the residual of some liquid after the salt drop slides over. This is borne out by further experiments. As shown in Fig. 3a and 3b, we can clearly see that there are many bright dots left behind and most of them are sitting on the step edges of the HOPG surface after a dilute salt drop slides over. In contrast, we cannot see dots left on the HOPG surface after a pure water drop slides over (Fig. 3c). It can be expected that the CAH can reach a maximum when sufficient cations are "blocked" by the stripe "dams" on the surface, showing the abrupt transitions of the CAHs in Fig. 1c.

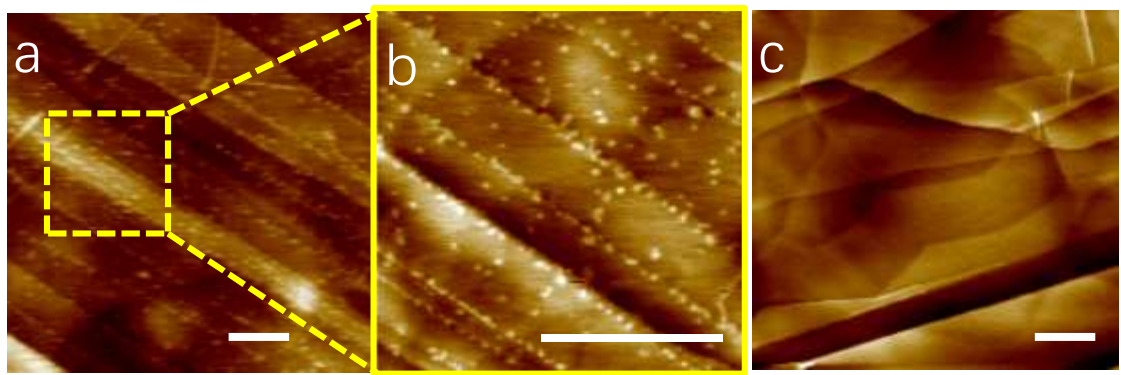

**Figure 3**. AFM height images of a HOPG substrate right after the sliding over of a drop (a) with $Na^+$ concentration of 0.25 M, and (c) of pure water; (b) Zoom-in image of the selected area in (a). Scan bars are 500 nm.

In summary, we experimentally observed the enhanced contact angle hysteresis (CAH) of dilute salt aqueous drops on graphite surface, which can reach 2.8-4.0 times the CAH of pure water and the explicit CAH value can be tuned by different cations. Theoretical studies show that the key to this unexpectedly experimental phenomenon is the strong cation-π interaction between cations and aromatic rings on the HOPG surface, which preferentially adsorbs cations on the surface. These adsorbed cations are blocked by the inherent atomic steps on the HOPG surface, hindering the sliding of the dilute salt aqueous drop like pins, which reduces the receding contact angle and hence enhances the CAH. It is worth noting that the unique cation-π interaction exists in the other systems with carbon aromatic rings, including some polymers (polyethylene terephthalate, and polyimide), biomolecules, humus in soil, and even airborne pollutants. [32] Our study also gives a feasible method to tune the CAH on these graphitic surfaces by changing ion species or controlling the concentration of dilute salt solution, and will open simple, efficient, and environment-friendly avenues for improving the industrial, agricultural, and environmental applications of graphitic surfaces.

**Acknowledgement**
We thank Yi Gao, Beien Zhu, Rongzheng Wan, Chunlei Wang, and Yusong Tu for their constructive suggestions and helpful discussions. We also thank the Jihao Li, Bowu Zhang, and Ziqiang Cheng for providing technical supports for the contact angle measurement. This work was financially supported by the National Natural Science Foundation of China (Grants No. U1632135, U1932123, 11974366 ), the Key Research Program of the CAS (Grants No. KJZD-EW-M03), the Key Research Program of Frontier Sciences of the Chinese Academy of Sciences (Nos. QYZDJ-SSW-SLH053 and QYZDJ-SSW-SLH019), the Deepcomp7000 and ScGrid of the Supercomputing Center, the Computer Network Information Center of the CAS, the Special Program for Applied Research on SuperComputation of the NSFC-Guangdong Joint Fund (second phase), the Shanghai Supercomputer Center of China.